# Strong attentional modulation of V1/V2 activity implements a robust, contrast-invariant control mechanism for selective information processing

Lukas-Paul Rausch[1,4], Maik Schünemann[2,4], Eric Drebitz[1], Daniel Harnack[3], Udo A. Ernst[2,5], Andreas K. Kreiter[1,5]

[1]*Brain Research Institute, University of Bremen, 28359 Bremen, Germany*

[2]*Computational Neurophysics Lab, Institute for Theoretical Physics, University of Bremen, 28359 Bremen, Germany*

[3]*Robotics Innovation Center, German Research Center for Artificial Intelligence, 28359 Bremen, Germany*

[4,5]*These authors contributed equally*

**Abstract**

When selective attention is devoted to one of multiple stimuli within receptive fields of neurons in visual area V4, cells respond as if only the attended stimulus was present. The underlying neural mechanisms are still debated, but computational studies suggest that a small rate advantage for neural populations passing the attended signal to V4 suffices to establish such selective processing. We challenged this theory by pairing stimuli with different luminance contrasts, such that attention on a weak target stimulus would have to overcome a large activation difference to a strong distracter. In this situation we found unexpectedly large attentional target facilitation in macaque V1/V2 which far surpasses known magnitudes of attentional modulation. Target facilitation scales with contrast difference and combines with distracter suppression to achieve the required rate advantage. These effects can be explained by a contrast-independent attentional control mechanism with excitatory centre and suppressive surround targeting divisive normalization units.

# Introduction

Complex natural environments often provide a plethora of sensory information that usually exceeds the brain's processing capacity. To cope with this excessive amount of sensory information, the brain employs mechanisms of selective attention to pick out behaviorally relevant input. The need for such selection mechanisms is particularly evident and well-documented for processing visual information in the primate brain. Due to the convergence of cortico-cortical connections, individual neurons are responsive to increasing proportions of the visual field along successive processing stages (Gattass et al., 2005). Consequently, adjacent objects within the visual field that activate separate populations of neurons in upstream areas fall within the same, larger receptive field (RF) of the common postsynaptic neurons in downstream areas. The resulting competition of the efferent signals originating from upstream neurons for processing resources of receiving neurons is resolved by attentional mechanisms (Desimone & Duncan, 1995). For instance, neurons in mid-level areas like V4 or MT in the macaque monkey respond selectively to the attended stimulus when two stimuli are simultaneously present within their RFs (Moran & Desimone, 1985; Reynolds, Chelazzi, & Desimone, 1999; Treue & Maunsell, 1996). This indicates selective processing of behaviorally relevant input despite the simultaneous presence of irrelevant afferent signals.

How such flexible signal selection is implemented mechanistically at the neuronal level is still an open question. Recent modelling studies demonstrate that only a modest rate advantage for the neuronal population processing the attended stimulus as compared the populations processing the distracter in upstream areas is required to enable subsequent mechanisms to achieve selective processing and representation of the relevant stimulus-information further downstream in the visual pathways (Harnack, Ernst, & Pawelzik, 2015; Palmigiano, Geisel, Wolf, & Battaglia, 2017). Indeed, it appears that shifts in spatial attention are associated with modulations of neuronal responses which establish such modest rate advantages, thus supporting this theory. Most neurons respond more strongly to a single stimulus shown in their RF when spatial attention is focused onto that stimulus compared to when attention is directed to a stimulus at a distant location (McAdams & Maunsell, 1999; Mountcastle, Motter, Steinmetz, & Sestokas, 1987; Reynolds et al., 1999; Spitzer, Desimone, & Moran, 1988; Treue & Maunsell, 1996; Williford & Maunsell, 2006). Furthermore, the same neurons respond less if attention is directed to a stimulus in the near surround of their RFs (Sundberg, Mitchell, & Reynolds, 2009). If attended (target) and unattended (distracter) stimuli elicit similar responses without any attentional bias, these attentional modulations reliably result in in typically modest rate advantages of total population activity for upstream populations that represent the target over those that represent the distracter (Chen et al., 2008; Roelfsema, Lamme, & Spekreijse, 1998; Thiele, Pooresmaeili, Delicato, Herrero, & Roelfsema, 2009). Well in line with selective processing of the stimulus evoking the higher population activity, also bottom-up influences like stimulus differences in contrast, color or size can bias selective processing. For example, a single outstanding high-contrast stimulus results in greater population responses than an otherwise identical low-contrast stimulus (Albrecht & Hamilton, 1982), and captures attention (Theeuwes, 1992). Taken together, these observations raise the question how the visual system would handle situations in which an attentional bias is competing with a bottom-up bias, for example when pairing a weak contrast stimulus with high contrast distracter.

Despite the expected large stimulus-induced activation advantage for the stronger stimulus, selective attention tasks are performed successfully also for the weaker stimulus (Mounts & Gavett, 2004; Pashler, Dobkins, & Huang, 2004), indicating that selective processing of a low contrast target amongst high contrast distracters is still possible.

According to modeling studies that propose a rate advantage as a prerequisite for successful selective information processing at downstream populations, these considerations imply that attention must induce large rate changes to achieve at least small activity advantages of the upstream population processing a low contrast target (Harnack et al., 2015). This situation becomes particularly challenging when a stimulus-driven imbalance is induced, which exceeds the (modest) magnitude of attentional modulations usually observed in areas V1 and V2.

Here we test this prediction in areas V1/V2 of macaque monkeys performing a demanding shape-tracking task that required sustained attention for one of several stimuli. We investigated to which extent attention-dependent rate modulations compensate for large contrast-dependent activity differences between two closely spaced, competing stimuli. Our results show that attentional rate modulations consistently compensate large stimulus-induced activity deficits and establish a rate advantage for the neuronal population encoding the target over those responding to the distracter stimulus. This occurs mainly by facilitating the target responses (65% average rate gain) and simultaneously, to a lesser extent, by suppressing the distracter responses (16.8% average rate attenuation). The rate advantage for V1/V2 populations representing the target was not achieved in error trials. Our findings reveal a surprisingly strong compensation ability for attention-dependent rate modulations in V1/V2, thus supporting models of a control mechanism where spatial attention elevates the population rate for the attended stimulus above that for the unattended stimuli to enable subsequent mechanisms to achieve selective processing of the signals encoding the behaviorally relevant object. We further show that a simple model can quantitatively reproduce attention-dependent rate modulations among all tested stimulus configurations. The model postulates a contrast-independent control mechanism for attention which has an excitatory center and a suppressive surround, and which in combination with lateral inhibition by stimulus context and a non-linear neural contrast-response function is able to explain the observed modulation effects.

## Materials and Methods

*Surgical Preparation*

Two male macaque monkeys (Macaca mulatta) were implanted with a titanium head post and a recording chamber over the prelunate gyrus under general anaesthesia and strictly aseptic conditions. Stereotactic coordinates of visual target areas were derived from anatomical landmarks and structural magnetic resonance imaging (MRI) scans before surgery.

All procedures were approved by the local authorities (Die Senatorin für Gesundheit, Bremen, Germany) and were in accordance with the regulation for the welfare of experimental animals issued by the Federal Government of Germany and with the guidelines of the European Union (2010/ 63/EU) for care and use of laboratory animals.

*Behavioral Task*

The monkeys performed a demanding shape-tracking task, which is a delayed match-to-sample task that requires the allocation of covert visual selective attention to a cued spatial location. During the inter-trial period (2000 ms), a small low-luminance annulus (1° diameter, 0.04° line width) was cueing the position of the upcoming behaviorally relevant stimulus (target). The animal initiated a trial by fixating on a central, rectangular fixation point (FP, 0.15° x 0.15°, 2.45 Cd/m$^2$) and pressing a lever inside the primate chair. Fixation had to be maintained throughout the whole trial within a circular fixation window of 1° diameter around the FP. After a baseline period of 1050 ms with a blank screen, three to four complex-shaped contour stimuli were statically presented on the screen for 520 ms. One or two stimuli (1.5° diameter, 0.25° line width) were presented in the lower visual quadrant that was contra-lateral to the recording sites in V1/V2. Here, one stimulus was placed inside the receptive field (RF) of the recorded neuronal population. The location of the other stimulus was chosen to be in close vicinity to the RF stimulus while avoiding to induce cross-activation. Two other stimuli were positioned point-mirrored in the upper quadrant of the visual field (Fig. 1A). After the static period, all stimuli started a continuous morphing process (Fig. 1C) where each stimulus morphed through a sequence of shapes as described in (Taylor et al., 2005). The reappearance of the initial shape (target) at the cued location had to be detected by the monkey and responded to by releasing the lever (510 ms response window starting 310 ms before target onset). Each trial comprised two to four sequences (morph cycles, MCs) with each MC lasting 1000 ms. The stimulus shapes for each sequence and stream were selected randomly from a set of 11 shapes (six for monkey B), allowing for the target shape of each trial to appear also in the distracter streams. Note that for monkey T all shapes out of the whole set could be chosen as the initial target shape, whereas only two could become the target shape for monkey B. Each of the four stimulus locations could be cued to constitute the target stream, resulting in different attentional conditions for the recorded population of neurons (Fig. 1B). Trials with attention allocated to a stimulus in the ipsilateral upper quadrant are denoted as attend-away trials (Fig. 1B, upper row). When attention was directed onto the stimulus inside the RF (attend-in, Fig. 1B lower left), the stimulus in close vicinity represented a nearby distracter. Likewise, the stimulus inside the RF constituted the nearby distracter when attention was cued to the adjacent stimulus position (attend-nearby, Fig. 1B lower right). Trials were terminated without reward if the monkey broke fixation or responded with premature or delayed lever releases.

Since neurons in V1 and V2 exhibit a monotonic contrast tuning (Albrecht & Hamilton, 1982), population activity was modulated by using stimuli of different contrasts. In particular, we employed stimulus configurations where the receptive field stimulus was accompanied by a stimulus of matching (low-low, high-high) or different luminance contrast (low-high) in the near surround (center-to-center distance of 1.5 – 2° visual angle, see Fig. 1B). The low and high contrast values were chosen based on the previously measured contrast response function (see section Electrophysiological recordings) recorded before the main experiment, with the requirement to induce an activity difference between the responses to low and high contrast stimulus that is as large as possible, while ensuring sufficiently high performance levels of the monkeys. Low contrast values ranged from 4-64% and high contrast values from 8-72%. Stimuli in the ipsilateral upper quadrant of the visual field had a fixed contrast of 50% over all conditions. By switching the contrast identity of the stimuli in the lower visual quadrant in the low-high stimulus configuration across all attentional conditions and assuming a similar activation by the stimuli across all populations we were able to investigate how attention modulates the neuronal activity for closely spaced target and distracter stimuli.

During the recording process it turned out that monkey T has myopic vision. As a consequence, he received refractive lenses for corrected vision (29 sessions without and 9 sessions with lenses). The myopia did only affect the psychophysical threshold for contrast perception but not the attentional effects.

### *Data Acquisition*

*Visual Stimulation*

Complex-shaped visual stimuli were generated with a custom-made software (VStim) and presented on a 22-inch CRT-monitor (1024 x 786 pixels, 100 Hz refresh rate). In the first stage (24 sessions Monkey T, 6 sessions Monkey B) of the recording process the Michelson-contrast was determined by the luminance of the shape and the background luminance of screen, which was set to 0.5 Cd/m$^2$. In a later stage, the shapes consisted of a one bright and one dark stripe which determined the stimulus contrast and deviated by the same absolute value from background luminance (Fig. 1A). In this case, the background luminance of the monitor was set to 5 Cd/m². The monkeys sat in custom-made vertical primate chairs with a distance of 95 cm (93 cm for monkey B) to the stimulation monitor. Eye-positions were tracked by video-oculography (IScan Inc., Woburn, MA, USA) with a sampling rate of 100 Hz.

*Electrophysiological recordings*

Intra-cortical recordings were performed in area V1 and V2 of two macaque monkeys using semi-chronically implanted glass-insulated tungsten microelectrodes (125 µm shank-diameter, 1 MΩ impedance, FHC Inc., Bowdoin, ME USA). The recorded signals were amplified 20000-fold for monkey T and 8000-fold for monkey B (4x by a wide-band preamplifier MPA32I, 5000x for monkey T / 2000x for monkey B by a PGA 64), 1-5000 Hz, both Multi Channel Systems GmbH, Reutlingen, Germany) and digitized by 12-bit AD converter with a sampling rate of 25 kHz. The neurophysiological signals were referenced against the titanium recording chamber, which enclosed the cortical areas of interest and made contact to the bone and dura mater. In the beginning of each recording session, the receptive field boundaries of the probed neuronal population were established by a manual mapping with appropriately sized bars (ca. 1°) of various orientations by reference to the audio-monitored multi-unit activity (MUA, 0.3-5 kHz). Considerable increase in MUA in response to the moving bar stimulus served as indicator for neuronal activation. During this procedure, the monkey was required to maintain fixation onto a small fixation point (0.15° x 0.15°) in the center of the screen. Subsequently, the contrast response function of the recorded population of neurons was assessed by presenting only a single morphing shape stimulus in their receptive field (without a second nearby stimulus) and varying its luminance contrast over trials in a pseudo-randomized order. While the monkey performed the attend-away condition of the task, the responses to stimuli of seven different luminance contrasts (ranging from 4% to 72%) were measured over approximately 50 trials resulting in about 14 MCs per luminance contrast.

*Confirmation of recording sites*

The implanted recording chambers provided access to area V1 and V2 in both monkeys. Position of recording sites were determined based on conventional retinotopic mapping relative to the representation of the vertical meridian (Gattass et al., 1981) and assessing estimated receptive field sizes based on an automated bar-mapping procedure described by (Drebitz et al., 2019). In addition, the polarity together with the latency of onset-transients gave supporting information for the verification of electrode position in the visual cortex.

*Spike thresholding and firing rate estimation*

Spiking activity was obtained by applying a high-pass equiripple FIR filter (300 Hz cutoff) in forward and backward direction to the recorded signal and using the thresholding method according to Quian-Quiroga (Quiroga, 2004):

$$thr = k \cdot median\left(\frac{|x|}{0.6745}\right)$$

The median of the bandpass-filtered signal was estimated over the baseline period of each individual trial and the scaling factor k was chosen manually for each recording session in the range from 3 to 7 with a median value of 3.5. Reasonable changes of k did only change the mean rates but did not affect attentional effects.

A time period of 150 ms before the behavioral response was dropped from analysis.

For quantitative analyses, we report mean firing rates over a 600 ms window (300-900 ms into the morph cycles) of the morph cycles 2 and 3 in hit trials. Statistical significance of rate differences resulting from different task conditions was tested using the Wilcoxon Signed Rank Test (WSR) on the corresponding mean firing rates of the recording sessions. To assess statistical significance of rate differences within a single session, the exact Wilcoxon Rank Sum Test (WRS) with continuity correction was used on the mean firing rates of the corresponding trials. The significance of attention-induced differences in the neural activity during the baseline period was tested with the Wilcoxon signed rank test on the pooled mean baseline firing rates of all task conditions with the respective attentional locus (-away, -in, -nearby). Because the attentional modulations of firing rates were qualitatively similar for both V1 and V2 and across changes in visual stimulation, we report the mean firing rates pooled across all recording sessions.

*Linear regression*

Linear regressions analysis was performed using a robust linear regression model with a bi-square weight function to reduce the impact of potential regression outliers (see *robustfit* implementation in MATLAB). This algorithm uses iteratively reweighted least squares to estimate model coefficients (Holland & Welsch, 1977). The tuning constant of the weight function was set to 4.685.

*Session inclusion criteria*

Recording sessions were included for further analysis if the overall performance (accuracy) of the monkey exceeded 66%, and if the cross-activation by a single, attended high-contrast stimulus nearby the RF with no stimulus inside the RF did not increase baseline activity by more than 15% of the response to a single, attended high-contrast stimulus inside the RF. Since we want to investigate the compensation of contrast-dependent activity differences by attentional rate modulations, we only considered sessions where the activity for the high contrast stimulus was at least 10% higher than for the low contrast stimulus in the attend away condition (corresponding to a minimum relative rate difference of 0.1). Furthermore, to allow for sufficient statistics, we required a minimum of 10 morph cycles from successfully terminated trials in the stimulus configuration where a high-contrast distracter flanked a low-contrast target.

*Model fits*

In the experiment, a suitable low and high contrast was chosen for each recording site, and 14 different combinations of stimulus contrast configuration and attentional condition were recorded. These included (a) four conditions obtained from combining single stimuli of low or high contrast with attention directed inside or away from the neuron's RF. Furthermore, there were (b) six conditions obtained from combining two stimuli of same contrast (configurations low-low or high-high) with three attentional conditions (away, in, and nearby). The remaining four conditions were obtained from pairing a low contrast stimulus with a nearby high contrast stimulus (and vice versa), in combination with attention either directed away or to the low contrast stimulus. The mean neural responses (baseline subtracted firing rates) arising in these 14 conditions comprised the target data for the model fits.

The six parameters $R_{max}$, $\sigma$, $\nu$, $b$, $a_{iI}$, $a_N$ of the model (Eq. (1)) were fitted to the experimental data using a non-linear least squares fit (Python library 'scipy.optimize') with the constraints $b \in (-1, 1)$, $a_I \in (-1, 1)$, $R_{max} > 0$, $\sigma > 0$, and $\nu > 0$. For simplicity, input drives $S_i$ and $S_n$ were set to the Michaelson contrast $C$ of the corresponding (low- or high-contrast) stimuli, with $S_i = 0$ or $S_n = 0$ marking the absence of a stimulus in the respective position. When attention was directed away, both $a_i = a_n = 0$. Since we allowed also negative values for $b$, $a_I$, and $a_N$, we asserted that both the numerator and denominator in Eq. (1) were non-negative for all conditions. In order to reproduce the large effect sizes of monkey T, deviations of the model responses for conditions in which a low contrast stimulus is paired with a second nearby stimulus were weighted by a factor of 5 instead of 1. For individual fits of each recording site, we took the best fit obtained from 100 runs with randomly chosen initial parameter values.

When fitting the model to responses averaged over all recording sites, the mean of the chosen low and high contrast values were used as model inputs ($S_{low} = 0.32$, $S_{high} = 0.63$ for monkey T and $S_{low} = 0.2$, $S_{high} = 0.34$ for monkey B). In some fits, the additional constraint $n = 1$ was imposed leading to a model with only five free parameters.

Quality of the model fits was assessed by the percent of variance (PVE) of the experimental activity that is explained by the model responses over the $N = 14$ configurations. With $R_i^{(e)}$ and $R_i^{(m)}$ being the experimentally observed response and the response of the chosen model for the configuration, respectively, PVE is defined as

$$PVE = 1 - \frac{\sum_{i=1}^{N}\left(R_i^{(m)} - R_i^{(e)}\right)^2}{\sum_{i=1}^{N}\left(\overline{R}_i^{(e)} - R_i^{(e)}\right)^2}$$

where $\overline{R}_i^{(e)} = \frac{1}{N}\sum_{i=1}^{N} R_i^{(e)}$ is the mean response.

*Data and Code accessibility*

Offline analyses and procedures were performed using custom made MATLAB (MathWorks, Natick, MA, USA) and Python software using the NumPy (Van Der Walt et al., 2011), SciPy (Jones et al., 2001), Matplotlib (Hunter, 2007) and pandas (McKinney, 2010) libraries. The data sets and software code generated during the current study are available from the authors on reasonable request.

**Results**

We recorded multi-unit spiking activity in visual areas V1 and V2 of two macaque monkeys performing a demanding shape-tracking task (Grothe et al. 2012, Taylor et al. 2005) requiring the animals to devote sustained covert attention to one of four stimuli (Fig. 1). These stimuli consisted of different shapes (size about 1°) which after an initial static period began to morph continuously into other shapes. For the stimulus at the previously cued location, the monkeys had to signal the reappearance of the initial shape during the morphing sequence after two to four morph cycles by releasing a lever. One of the stimuli was placed in the recording sites' RF (in the contralateral lower quadrant of the visual field). A second stimulus could appear at an isoexcentric location just outside the RF but close enough, such that both stimuli would fall into the larger RF of their common downstream neurons in areas like V4 or MT where they compete for being processed. Two further stimuli occurred in isoexcentric positions in the ipsilateral, upper quadrant, allowing to direct attention far away from the recorded neurons' RF.

For quantifying neuronal activity, we estimated the mean firing rate within 600 ms windows covering the behaviorally relevant periods (between 300 – 900 ms) in morph cycles 2 and 3 (unless mentioned otherwise). In total, 53 recording sites (Monkey T: 38 sites, Monkey B: 15 sites) fulfilled the inclusion criteria (see Methods), that ensured a sufficient number of trials and satisfying performance levels, and were used in the subsequent analyses. Firing rates in the baseline period before stimulus onset were not significantly modulated by attention in any of the attentional conditions (Wilcoxon signed-rank test: attend-away versus attend-in: $p = 0.9$ (T: $p = 0.88$, B: $p = 0.6$); attend-away versus attend-nearby: $p = 0.504$ (T: $p = 0.88$, B: $p = 0.06$); attend-in versus attend-nearby: $p = 0.92$ (T: $p = 0.53$, B: $p = 0.08$)

*Attention reverses even strong stimulus-induced activity differences between weak target- and strong distracter stimuli*

Presenting two closely spaced stimuli with large differences in contrast that were close to the monkeys' limits for performing the task, resulted without attention (attend away condition) in firing rates being on average 83.4% larger for the high as compared to the low contrast stimulus (Fig. 2A, T: 94.8%, B: 51.9%). This raises the question whether attention-dependent mechanisms are capable to overcome such substantial stimulus-dependent rate differences in areas V1 and V2 to establish a rate advantage for the neuronal population that represents the attended stimulus. Accordingly, we compared the mean firing rates evoked by the low contrast target (attend in, Fig. 2B) to the rates of the nearby high contrast distracter (attend nearby). Indeed, we find that attentional modulations (equilibrate or) reverse this activity imbalance in favor of the low contrast target stimulus so that the target activity matches distracter activity in 10 out of 53 recording sites, and significantly exceeds distracter activity in 38 out of 53 recording sites (T: 27, B: 11). Only 5 sites showed significantly higher mean firing rates for the high contrast distracter (Wilcoxon signed-rank test, $p < 0.05$).

Considering the activity of the ensemble of recorded sites as a representative measure for neuronal population activity, the attended low contrast stimulus evokes a significantly higher population activity (mean: 47.9 spikes/s, T: 52.1 spikes/s, B: 37.4 spikes/s) than the nearby high contrast distracter (mean: 44.3 spikes/s, T: 48.1 spikes/s, B: 34.9 spikes/s, Wilcoxon signed-rank test, $p < 0.01$, $p < 0.01$, $p = 0.15$) resulting in a mean rate advantage for the target stimulus of 8.1% (T: 8.4%, B: 7.3%).

In the stimulus configuration where the competing stimuli have matching contrasts, both stimuli are expected to induce equal levels of neuronal activity when attention is directed far away. Accordingly, only little attentional firing rate modulations are required to establish a firing rate advantage for the target stimulus. Since the contrast values for low (LL) and high (HH) contrast stimuli varied between recordings sessions, resulting in two strongly overlapping populations of contrasts, we pooled the mean firing rates for the LL and HH stimulus configurations in the subsequent analyses. Comparing the neuronal activity for the attended stimulus to the activity for the nearby distracter stimulus shows that firing rates in response to the target are consistently exceeding those in response to the distracter stimulus (Fig. 2C). In contrast to the stimulus configuration with non-matching contrast the rate advantage for the target stimulus is much stronger (43.2% mean rate advantage, T: 44.6%, B: 38.7%). Interestingly, attention-dependent facilitation in the matching contrast condition was lower than predicted from the facilitation effect observed in the non-matching condition (Fig. 2C & D, grey dots), demonstrating that facilitation is adjusted to the stimulus contrast configuration.

*Target facilitation scales with firing rate difference*

Large firing rate differences of on average 83.4% between neuronal activity evoked by low and high contrast stimuli require large attentional modulations to achieve the observed rate advantage in favor of the low contrast target. This can be achieved by changes of the activity evoked by the target stimulus as well as by changes of the activity caused by the distracter stimulus, or by a combination of these two effects. We therefore disentangled how attention altered neuronal activity in response to the target and to the distracter stimulus. To compare modulation effects between recording sessions, we normalized all activities to the activity in response to the low contrast stimulus flanked by a high contrast stimulus without attention (LH, attend away condition).

Directing attention onto the low contrast increased the activity of neurons processing the target stimulus consistently across sessions, reaching rate enhancements of more than 200%. Since the observed large variability in rate difference between low and high contrast stimuli challenges putative mechanisms for establishing a rate advantage to varying degrees, we analysed how target facilitation depends on the rate difference between low and high contrast stimulus in the attend away condition.

The linear regression analysis revealed a strong and significant correlation between the attention-dependent increase in activity for the low contrast stimulus (target facilitation) and the relative rate difference (($FR_{HL}$ – $FR_{LH}$)/$FR_{LH}$, Fig. 3A, r = 0.91, p = 2.25e-21, T: r = 0.82, p = 4.17e-15, B: 0.72, p = 3.6e-5), showing that the strength of target facilitation scales with the size of the difference. This observed correlation may be a spurious reflection of the strength of the low contrast stimulus. If attention-dependent additive rate enhancement were the cause, we would expect larger relative rate changes for lower firing rates and larger relative rate differences for distracter stimuli approaching saturation levels. However, our findings reveal a wide and uniform spread of attention-dependent increases in firing rates across all activity levels (standard deviation = 15.5 spikes/s). Moreover, we did not observe a significant negative correlation (r = 0.36, p = 0.07) between the relative target facilitation and stimulus-induced activity, which would be expected if a constant rate enhancement were present. The slope of the linear fit (y = 0.73x + 0.03, T: 0.83x – 0.08, B: 0.36x + 0.21) indicates that the attentional rate gain already compensates for about 73% of the relative rate difference. However, in most cases (40 out of 53) target facilitation is not sufficient to establish a rate advantage over the neuronal activity evoked by the nearby high contrast distracter with attention directed far away. Consequently, the activity evoked by the high contrast stimulus must have been attenuated to yield the observed rate advantage for the target. Analysis of this distracter suppression will be covered in detail in a later section, but combing the rate modulations for the target and distracter stimulus indeed results in sufficient compensation of the rate difference. As shown in Fig. 3B the cumulative attention effect is strongly correlated with the relative rate difference (r = 0.95, p = 3.02e-27, T: r = 0.9, p = 2.1e-19, B: r = 0.94, p = 2.54-09). The linear fit shows that the combined attentional modulations scale with the rate difference by about 1.14 with an offset of 0.06 (T: 1.2x – 0.07, B: 0.85x + 0.3).

*Competing stimuli of matching contrast induce weaker facilitation effects*

We already showed that the attended stimulus is associated with a firing rate advantage over the nearby distracter stimulus for stimulus configurations of non-matching and matching contrast (Fig. 2B & C). In the configuration with matching contrast, a scenario that is typically used in visual attention tasks, the stimuli are assumed to induce similar neuronal activity. Consequently, little attentional modulation is already sufficient to establish the rate advantage for the target stimulus because there is no rate difference that needs to be compensated. However, different contrasts are ultimately associated with different levels of stimulus-induced activity and the range of different contrast over all recording sessions enabled us to investigate potential activity-dependent effects of attentional modulations.

Analysis of activity-dependent target facilitation for the matching contrast stimulus configurations reveals that the attention-dependent gain in spiking activity does not show any significant correlation with the stimulus-induced neuronal activity without attention (Fig. 3C, r = 0.25, p = 0.13, T: r = 0.17, p = 0.45, B: r = 0.19, p = 0.44). In fact, the absolute changes in firing rates due to attention vary strongly across activity levels from the non-attended condition. For a mean activity of 43.9 spikes/s (T: 48.2 spikes/s, B: 32.9 spikes/s), we observed a mean increase in firing rates associated with attention of 9.9 spikes/s (22.7%, T: 21.1%, B: 28.2%). These results indicate that the increase in firing rates for the target stimulus in such stimulus configurations may be best described by an additive rate enhancement.

We also quantified the size of attention effects in the stimulus configuration with matching contrast by computing the classical attentional modulation index (AMI) for each recording site and stimulus identity (target and distracter, Fig. 3D). The AMI is defined as the difference in mean response rates between the attend in (attend nearby) and attend away condition divided by the summed responses. We find a positive shift of the AMI distribution for the target stimulus with a mean value of 0.16 ($p < 0.01$, t-test, T: 0.17, $p = 6.6e-09$, B: 0.16, $p = 7.66e-10$). In contrast, we observed a shift of the modulation indices to negative values for the nearby distracter stimulus with a mean value of -0.14 ($p < 0.01$, t-test, T: -0.17, $p = 3.44e-07$, B: -0.07, $p = 0.0035$).

*Distracter suppression is highly similar across stimulus configurations*

When attention was directed to the stimulus just outside the classical receptive field, we observed a decrease in firing rates in response to the distracter stimulus within the receptive field, regardless of whether the stimulus contrast matched or did not match the target. For target and distracter pairs of matching contrast we found a significant correlation between the distracter suppression and the activity evoked by the same stimulus configuration without attention (Fig. 4A: $r = 0.62$, $p = 1.28e-06$, T: $r = 0.55$, $p = 3.2e-4$, B: $r = 0.72$, $p = 0.0044$).

Interestingly, we observed a similar activity-dependent pattern of distracter suppression in the stimulus configuration with matching stimulus contrast (Fig. 4B: $r = 0.62$, $p = 8.24e-12$, T: $r = 0.63$, $p = 6.28e-09$, B: $r = 0.4$, $p = 0.053$). The similarity of regression coefficients suggests that the attention-dependent suppression of distracter activity is independent of stimulus configurations and the potential rate differences between competing stimuli, but instead relies on the activity level evoked by the distracter stimulus. Accordingly, we found that the average attention-dependent distracter suppression was consistent, regardless of whether the target had a weaker or equal contrast. In the non-matching contrast stimulus configuration, with an average firing rate of 53.3 spikes/s (T: 58 spikes/s, B: 41.3 spikes/s), we observed an average suppression of firing rates of 16.8% (T: 17.1%, B: 15.6%). Similarly, for competing stimuli of matching contrast, with an average firing rate of 43.5 spikes/s (T: 48.2 spikes/s, B: 32.9 spikes/s), the activity associated with the distracter was suppressed by 14.3% (T: 16.3%, B: 7.5%). These findings suggest that the attention-dependent suppression of distracter activity is consistent across different stimulus contrasts and primarily depends on the level of activity evoked by the distracter stimulus.

*Incomplete compensation of rate difference in error trials*

To examine the relationship between attention-dependent rate modulations and behavior, we analyzed the firing rates in error trials, which encompassed premature (false alarm) or omitted (miss) lever releases. If a firing rate advantage for the attended stimulus is crucial for successful information processing in downstream populations, a partial compensation of the initial rate imbalance that fails to achieve this advantage would be associated with task failure.In the stimulus configuration with non-matching contrast, we observed a higher proportion of recording sites that exhibit higher distracter-related activity than target-related activity (Fig. 5A, points below diagonal). The resulting shift of the population mean below the identity line indicates that attentional modulations do not fully compensate for stimulus-induced rate differences. The mean firing rate of multi-uni activity (MUA) evoked by the high contrast distracter (46.8 spikes/s) remained larger than the MUA associated with the low contrast target (42.7 spikes/s) resulting in a mean rate disadvantage of 9.6% (T: 5.1%, B: 26%) for the target. However, the population activity difference is not significant (Wilcoxon signed-rank test, p = 0.11, T: p = 0.36, B: p = 0.09). Only 41% (T: 46%, B: 33.3%) of the recording sites still attain a rate advantage for the target stimulus. For correctly executed trials, these were 71.7% (T: 71%, B: 73.3%).

The comparison of mean firing rates for target and distracter stimuli of matching contrasts draws a different picture (Fig. 5B). Although we find an increased number of recording sites with a rate advantage for the distracter stimulus, the majority of recording sites (77.8 %) still exhibit higher activity levels for the target stimulus in error trials. Compared to the successfully conducted trials, the rate advantage of the population mean dropped from 43.15% to 30.1% (T: 38.4%, B:11%).

*A minimal model for response modulation by attention and stimulus configuration*

The observed rate advantage for a target is established by the combination of two effects, target facilitation and distracter suppression, with the former depending strongly on the relative activation difference caused by the stimulus configuration. We wondered if we could propose a minimalistic control mechanism for attention that suggests a physiological explanation for these main effects and reproduces neural response rates across all stimulation conditions.

For constructing a corresponding model, we make the standard assumption that the dependence of neural responses on stimulus drive (i.e., contrast) is well described by a non-linear, saturating contrast-response function (Sclar et al., 1990). Stimulus drive increases with the contrast of the stimulus inside the neuron's RF, while the presence of a second, nearby stimulus lowers this stimulus drive proportionally to its own contrast (surround suppression). We then hypothesize that attention acts on two parallel control mechanisms: Attention on the stimulus within the RF induces a constant shift of the contrast-response function to the left (contrast gain enhancement; Fig. 7A), while attention on a nearby stimulus causes divisive attenuation of the neural response for the distracter stimulus.

In combination with the non-linear contrast-response function, the magnitude of these two effects will then vary with stimulus contrast: while target facilitation will be larger when target contrast is low, distracter suppression will become stronger when distracter contrast is high (Fig. 7A). When a target is paired with a distracter, both effects combine and extend the range of target contrasts over which the target will have a rate advantage over the distracter (Fig. 7C). In particular, surround suppression kicks in and makes those effects dependent on the whole stimulus configuration.

Both mechanisms can be naturally implemented in a neuronal circuit (Fig. 7B) inspired by the well-known normalization model of attention (Reynolds & Heeger, 2009). The steady-state neural response $R_i$ to a stimulus configuration with one stimulus placed inside the neuron's classical receptive field (cRF) and a second stimulus placed at a nearby location (for examples see Fig. 1) is described by the following equations:

$$S_I = S_i - bS_n + a_i$$

$$R_i = g(S_i) = R_{max} \frac{S_I^v}{\sigma^v + S_I^v + a_n^v} \quad (1)$$

In these equations, $S_i$ and $S_n$ denote the input drive provided by the stimuli inside and nearby the neuron's cRF with contrasts $C_i$ and $C_n$, respectively. $a_i$ is a facilitating attentional control input with a value $a_I$ larger than zero if attention is on the stimulus inside the cRF (and zero otherwise). $a_n$ is a suppressive attentional control input that enhances divisive inhibition with a value $a_N$ larger than zero if attention is on a nearby stimulus (and zero otherwise). While $a_I$ and $a_N$ are fixed model parameters, $a_i$ and $a_n$ are variables that depend on the direction of attention. Consequently, $a_i = a_I$ and $a_n=0$ designate attention on the stimulus inside the RF (the situation exemplified in Fig. 7B), $a_i = 0$ and $a_n = a_N$ describe attention on the stimulus nearby, and $a_n = 0$ and $a_i = 0$ represent attention directed away. $a_i$ and $a_n$ can never be simultaneously different from zero, since attention was never directed to two stimuli in parallel. $S_I$ results from combining the input drive $S_i$, the surround inhibition due to $S_n$ scaled by the factor $b$, and the facilitatory top-down modulation $a_i$. Finally, $R_i$ is the output obtained by divisive inhibition of $S_I$ with the suppressive top-down modulation $a_n$ entering the denominator. $R_{max}$ serves as a scaling factor for matching the circuit's output $R_i$ to a spike rate, while $\sigma$ determines the half-maximal response (for $a_n=0$), and hence the position of the contrast-response curve along the contrast axis. $\nu$ is an (optional) exponent determining the exact shape of the input-output nonlinearity. In summary, Eq. (1) describes a contrast-response function $g$ which maps an input $S_i$ to an output $R_i$. Its exact shape is determined by the fixed parameters $b, \nu, \sigma, R$ as well as by the current stimulus context contained in $S_n$ and attentional state in $a_i, a_n$.

      The model integrates stimulus- and attention-related input in two stages: first, direct stimulus drive $S_i$, surround inhibition $bS_n$, and, if attention targets the stimulus inside the cRF, top-down modulation $a_i$ implementing target facilitation are integrated by simply summing them into $S_I$. The second stage implements the non-linear contrast-response function. In parallel, it realizes divisive normalization, and, if attention targets the stimulus nearby the cRF, top-down modulation $a_n$ realizing suppression of responses $R_i$ to the distracter.

In particular, the experimentally observed dependence of target facilitation on the difference between target and distractor-evoked responses without attention originates from the difference $S_i - bS_n$. Hence decreasing target contrast as well as increasing distracter contrast will reduce this term, and thus induce a shift on the transfer function to the left, which leads to a larger gain when attention kicks in ($a_i > 0$). Distracter suppression is subject to a similar effect, which, however, has a different characteristics since $a_n$ only enters the denominator, and which can also be very different in magnitude.

*Fit to population average.*

In the experiments, neural responses were recorded for 14 different combinations of stimulus configuration and attentional condition (details provided in Methods). For predicting the mean neural responses (i.e., the baseline-subtracted firing rates) over the recorded population, we chose $\nu = 1$ for simplicity and only adjusted the five parameters $R_{max}$, $\sigma$, $b$, $a_I$ and $a_N$ using a non-linear least-squares fit. The quality of the model fits was assessed by the percent of variance (PVE) over the mean neural responses that is explained by the model responses over the 14 combinations. For monkey B, the fit explained 93.1% of the variance with the parameter values $R_{max} = 64.37$ spikes/s, $\sigma = 0.34$, $b = 0.11$, $a_I = 0.16$ and $a_N = 0.09$, while for monkey T, the fit explained 86.7% of the variance with parameter values $R_{max} = 75.91$ spikes/s, $\sigma = 0.58$, $b = 0.36$, $a_I = 0.26$ and $a_N = 0.32$.

*Fit to individual recording sites.*

Given the good agreement between the model and mean experimental data, we wondered if we could also reproduce the variability of neural responses and effect sizes over different recording sessions. For this purpose, we performed individual fits for all 53 sites recorded in both monkeys and obtained a median PVE of 0.765 (0.768 for monkey T and 0.706 for monkey B). When comparing the fits to the experimental data shown in the scatter plots in Figures 2-4, we observed that the model captures both shape and extent of the neural response distributions very well (not shown). In particular, the model reproduces the corresponding effects of attention and stimulus context also quantitatively, as can be seen from the comparison of linear regression coefficients obtained for model and neural data in Table 1.

Median values for the fit parameters were $R_{max} = 85.92$ spikes/s, $\sigma = 0.765$, $\nu = 0.786$, $b = 0.204$, $a_I = 0.231$ and $a_N = 0.498$ ($R_{max} = 82.95$ spikes/s, $\sigma = 0.74$, $n = 1.63$, $b = 0.22$, $a_I = 0.22$ and $a_N = 0.61$ for monkey T, and $R_{max} = 100.65$ spikes/s, $\sigma = 1.07$, $\nu = 0.42$, $b = 0.15$, $a_I = 1.0$ and $a_N = 0.15$ for monkey B). The distribution of individual values for $b$ and $a_I$ is shown in Fig. 8(A). The attentional control parameters induced a median contrast gain of $a_I = 0.231$ (target facilitation), leading to a rate increase of 16.7 spikes/s (18.9 spikes/s for experimental data) in the LH condition, 12.0 spikes/s (10.9 spikes/s for experimental data) in the LL condition, and 8.0 spikes/s (8.7 spikes/s for experimental data) in the HH condition. On average, the distracter suppression parameter $a_n$ caused a reduction of the model response to two stimuli with matching contrasts by 8.3 spikes/s (8.7 spikes/s for experimental data) for high contrasts, and by 5.9 spikes/s (4.1 spikes/s for experimental data) for low contrasts. These attentional modulation sizes and their standard error of the mean are shown in Fig. 8(B).

Using the individual fits, we computed the mean model response over the recorded population and compared it to the average neural response. The model explained 96.94% of the variance in the experimental data for monkey B and 96.66% for monkey T. The dependence of attentional modulation on stimulus configuration is shown in Fig.

8(C, D) for both, model and neural data. In particular, it becomes obvious that a fixed control input ($a_I$ or $a_N$) in combination with a non-linear contrast-response function indeed suffices to explain why target facilitation is stronger for low-contrast targets (panel C), while distracter suppression becomes larger for high-contrast distracters (panel D). Furthermore, the similarity of data distributions demonstrates that not only the average effects, but also their variability are well captured by the model.

| Figure | intercept | | slope | | expl. var. $R^2$ | | size comparison $(\bar{y} - \bar{x})/\bar{x}$ | |
|---|---|---|---|---|---|---|---|---|
| | Exp | Model | Exp | Model | Exp | Model | Exp | Model |
| 2A | 12.9 | 12.0 | 0.305 | 0.339 | 0.334 | 0.441 | 0.834 | 0.769 |
| 2B | 7.56 | 6.11 | 0.937 | 0.923 | 0.825 | 0.864 | 0.0814 | 0.0359 |
| 2C | 7.4 | 9.24 | 1.19 | 1.19 | 0.775 | 0.832 | 0.431 | 0.471 |
| 3A | 0.0423 | 0.0854 | 0.741 | 0.643 | 0.799 | 0.75 | 0.128 | 0.171 |
| 3B | 0.0724 | 0.112 | 1.11 | 0.98 | 0.901 | 0.862 | 0.134 | 0.0754 |
| 3C | 11.6 | 7.17 | 0.087 (n.s.) | 0.0452 (n.s.) | 0.0611 | 0.0301 | 0.773 | 0.765 |
| 4A | -3.8 | 1.56 | 0.26 | 0.104 | 0.382 | 0.123 | 0.832 | 0.857 |
| 4B | -5.42 | -0.205 | 0.272 | 0.0157 | 0.387 | 0.261 | 0.858 | 0.84 |

Table 1: Regression parameters for experimental data (Exp) shown in the corresponding figures (left column), in comparison to regression performed on the individual site model fits (Model). From left to right, intercepand slope of regression (marked with n.s. if not significantly different from 0), explained variance $R^2$, and the size comparison $(\bar{y} - \bar{x})/\bar{x}$ between the average of the respective x and y coordinate values.

**Discussion**

In this study, we investigated the capacity of attention-dependent firing rate modulations in areas V1 and V2 of the macaque monkey to overcome differences in stimulus-driven activity between neuronal populations representing closely spaced, competing stimuli. In particular, we were interested whether attentional modulations overcome even large differences between the population activities evoked by a low-contrast target and a high-contrast distracter stimulus and consistently establish a rate advantage in favor of the attended stimulus. We found that spatial attention indeed over-compensates the stimulus-dependent activity imbalance between the neuronal populations over an extensive range of original rate differences observed without attention to the two stimuli (up to more than 200%).

This change resulted from the facilitation of activity evoked by the low-contrast target and suppression of activity evoked by the nearby high-contrast distracter. Most notably, the strength of attention-dependent facilitation of the target-evoked activity scales with the original rate difference and compensates for about 73% of that imbalance. On the other hand, the distracter suppression depends mainly on the original rate evoked by the distracter alone. Together both mechanisms resulted in a minor but highly consistent rate advantage for the attended low-contrast stimulus over the competing high-contrast distracter of about four spikes/s during task-relevant periods. In contrast to this advantage during successfully performed trials, there was no rate advantage observed for the attended low-contrast stimulus if the monkeys failed to respond correctly.

Finally, we sought a minimal neuronal model explaining the observed effects of attention and predicting the detailed pattern of response strengths in the different experimental conditions. This was achieved by a model combining target facilitation due to an additive contribution of attention to the stimulus input evoked by the target and distracter suppression by an attention-dependent enhancement of divisive inhibition in the visual field representation surrounding the focus of attention.

Together, target facilitation and distracter suppression created a rate advantage for the target over a large range of activity differences, but this rate advantage might be very small for demanding conditions. For low contrast targets, they improve the signal-to-noise ratio (SNR) to only about one if both distracter and target populations project with equal weights to a downstream (e.g., V4) population. While this might be an improvement as compared to the smaller SNR expected based on the contrast tuning alone, it is clearly not sufficient to explain that down-stream neurons respond predominantly as if only the attended stimulus is present (Desimone & Duncan, 1995; Ghose, 2009; Reynolds & Heeger, 2009; Treue & Martinez Trujillo, 1999), and that their activities carry an at least two times higher stimulus information content about the target (Grothe et al., 2018). Hence, selective attention needs a second mechanism capable of translating the small rate advantage to a large difference in effective signal routing. A possible candidate is differential phase-synchronization in the γ-band (Fries, 2005; Kreiter, 2006), for which experimental evidence and physiologically realistic model implementations are available. Such neural network models facilitate selective information transfer through selective γ-band phase synchronization between one of the two (or more) upstream populations activated by different stimuli and the common receiver population of downstream neurons. Concurrently they suppress the transmission of signals from the other upstream populations responding to currently irrelevant stimuli to the common receiver population. In particular, Harnack et al. have shown with their model that a small population rate advantage induced in one of two active V1/V2 populations suffices to establish such a γ-band phase synchronization that selectively facilitates signal transmission between this population and the receiving V4 population (Harnack et al., 2015). The model's mechanism that switches synchronization with the receiving V4 population between the different V1/V2 populations relies on the existence of a bi-stable state established by recurrent interactions, which allows a small population rate advantage to switch between states and thus to induce large effects on effective signal transmission. The results of the present study support the concept of a population-rate controlled routing-by-synchronization mechanism of attention by demonstrating that even in V1 and V2, attention results, on the one hand, in particularly strong rate modulations if necessary to achieve an, on the other hand, limited population rate advantage for the population processing the behaviorally relevant stimulus. Well in line with the postulated functional relevance of the firing rate advantage for selective stimulus processing, we found that behavioral failures were associated with the population rate for the target stimulus not attaining a rate advantage.

The rate advantage thought to control a routing-by-synchrony mechanism was observed in most recording sites, albeit not all. However, this observation does not conflict with the proposed control mechanism because the relevant mechanism implemented by the model of Harnack et al. (2015) relies on the ratio between population activities. In contrast, our recordings encompass the activities of the few neurons that a single microelectrode can record simultaneously. Therefore the neuronal activity measured at each recording site is just a subsample of the underlying population activity. In line with the proposed population-activity-based mechanism, formation of a pseudo-population (by taking the data of the recording site together) resulted in a significantly higher population rate for the target stimulus than for the nearby distracter stimulus.

Aside from population rate modulations, also other factors like differences in $\gamma$-band frequency have been suggested to control the $\gamma$-synchronization between a receiver population and the competing sender populations Bastos et al., 2015; Palmigiano et al., 2017; ter Wal & Tiesinga, 2017; Voloh et al., 2016). If, in turn, rate changes modulate these factors, they might contribute as a link between rate change and change of γ-synchronization or they might contribute in parallel to rate changes to a differential control of γ-synchronization. For example, theoretical and experimental studies suggest that increasing the activity of a local population also increases the γ-band frequency (Lowet et al., 2015). Such an enhancement of the γ-band frequency was, in turn, proposed as a mechanism to take over the γ-synchronization with a downstream target population from competing input populations with lower γ-band frequencies (Fries, 2015).

Our findings reveal for areas V1/V2 a surprisingly strong capacity for attention-dependent firing rate increases reaching more than 200%, provided that rate differences to the distracter-evoked activity were sufficiently large. Previous studies of attention-dependent response modulations in these areas commonly reported rather limited effects that rarely exceeded 10 to 20% (Haenny & Schiller, 1988; Hembrook-Short et al., 2017; McAdams & Reid, 2005; Ruff & Cohen, 2016)Haenny & Schiller, 1988; Hembrook-Short et al., 2017; McAdams & Reid, 2005; Ruff & Cohen, 2016Haenny & Schiller, 1988; Hembrook-Short et al., 2017; McAdams & Reid, 2005; Ruff & Cohen, 2016Haenny & Schiller, 1988; Hembrook-Short et al., 2017; McAdams & Reid, 2005; Ruff & Cohen, 2016). A likely reason for this discrepancy is the stimulus configurations used in these studies, which employ competing stimuli that evoke population activities of almost the same strength. Therefore only small attentional modulations are required to establish a firing rate advantage for the population processing the attended stimulus. Thus, the absence of significant stimulus-dependent rate differences between the populations that deliver competing input to downstream areas accounts for the lack of substantial attentional modulations in such paradigms. Accordingly, we also observe much weaker rate enhancements for the target stimulus in configurations with competing stimuli of similar or matching luminance contrast. The present study shows that these comparatively weak attentional modulations do not reflect any limitation of attention-dependent rate modulations in V1 or V2 in principle. At least in the case of a low-contrast target stimulus, the much weaker attention-dependent rate modulation in the presence of a neighboring low-contrast distracter as compared to a high-contrast distracter further supports the conclusion that the primary functional role of attention-dependent rate modulations is not the maximization of population rate differences between the corresponding neuronal populations. Instead, it supports a role in controlling selective signal routing and processing due to other mechanisms (as described above) by establishing moderate population rate advantages, no matter whether this requires weak or strong rate modulations.

Stimuli located in regions surrounding a neuron's receptive field can considerably modulate the responses to stimuli presented within the receptive field. These contextual modulations depend in passive stimulation paradigms on the relative differences in stimulus properties like orientation, direction of motion, and luminance contrast between the concurrent stimuli (Kasamatsu et al., 2010; Levitt & Lund, 1997; Polat Uri et al., 1998; Sillito & Jones, 1996). Under top-down influences like attention, the magnitude of the induced modulatory effects markedly depends on the stimulus configuration, showing stronger target facilitation in the presence of competing (Chen et al., 2008; Motter, 1993; Roelfsema et al., 1998; Verhoef & Maunsell, 2016) or collinear stimuli (Ito et al., 1998; Ito & Gilbert, 1999; Li et al., 2006). In our experiment, the configuration of stimuli was constant over the recording sessions, with only the contrast of the receptive field stimulus and its proximal adjacent stimulus being changed, resulting in different contrast contexts for the target stimulus. In this scenario, we find that the attention-dependent rate gain of the attended stimulus strongly depends on the contrast-mediated rate difference to the adjacent stimulus

Target facilitation was always accompanied by distracter suppression, which is in line with earlier findings where attention to a stimulus outside the cRF reduces the response to a non-attended stimulus inside the cRF (Ito & Gilbert, 1999; Sundberg et al., 2009; Verhoef & Maunsell, 2016). In contrast to the scaling of target facilitation with the relative rate difference, the suppressive effect of attention on the nearby distracter increased with the activity evoked by that stimulus in the attend-away condition. Due to the monotonic nature of the contrast-response curve of neurons in V1 (Albrecht & Hamilton, 1982; Tang et al., 2021), this implies that attention-dependent suppression of the distracter-related activity mainly depends on the distracter's own stimulus contrast. Such activity-dependent surround-suppression likely arises from increased excitation of inhibitory neurons through interareal-feedback connections (Angelucci et al., 2017; Boehler et al., 2009; Nassi et al., 2013). We model this interaction by divisive inhibition, which provides a natural explanation for why suppression is more prominent for higher distracter contrasts. In addition, this effect might provide a metabolically beneficial mechanism for target selection. Stimuli of high contrast induce firing rates close to or at the saturation level of the contrast response curve. Particularly in this scenario, it is more cost-efficient to more strongly attenuate the responses to potentially multiple and less relevant distracter stimuli, and to less excessively elevate target-related activity to establish the target stimulus's rate advantage.

The surprisingly strong, attention-dependent modulations of neuronal activity and their critical dependence on the stimulus context could be explained by a model based on biologically plausible assumptions, even in quantitative details. Its network structure and interactions are compatible with visual cortical circuits' physiological and anatomical properties, while parameter distributions cluster in physiologically meaningful regions, with, e.g., surround inhibition being lower than direct excitation and target facilitation inducing a contrast shift of up to 50% (excluding some extreme cases).

Furthermore, the model provides a consistent quantitative explanation for all measured experimental conditions (detailed in Table 1). It allowed for good fits even for the individual recording sites. This suggests that although different neural populations might differ in parameters, such as maximum firing rate or effectiveness of attentional modulation, the model implements a functional principle that underlies the attention- and context-dependent response pattern across recording sites.

A potential objection could be raised against the ring-like structure of the surround, which might imply anatomically more demanding connection schemes than a disc-like structure, as in the Difference of Gaussian models. Therefore, we computed the fits for the individual recording sites to a model with the summand $a_n^\nu$ in the denominator of both, the equation describing responses $R_n$ of neurons in the suppressive surround and of the equation describing responses $R_i$ of neurons in the facilitatory center. Comparison of the mean model response based on these fits over the recorded population to the average neural responses revealed a strong match between model and experimental data (PVE of 93.71% and 93.72% for monkeys B and T, respectively). The similar quality of the fits indicates that our model does not critically depend on the ring-like structure of the surround.

Structurally our model explains the experimental observations by proposing the existence of two different mechanisms. The first realizes an attentional 'spotlight' with an excitatory center that increases the contrast-gain of neurons responding to the attended stimulus. The second reduces the output gain of neurons responding to distracter stimuli within a suppressive surround by divisive inhibition. This choice was strongly motivated by the observation that distracter suppression and target facilitation scaled qualitatively differently with increasing stimulus contrasts and contrast differences. Fits of model variants with only one of these mechanisms were much less successful (results not shown). Hence postulating a single control mechanism instead would require identifying a yet unknown, recurrent, non-linear dynamic capable of yielding attentional modulations with opposing effects and different dependence on stimulus contrast.

In summary, the results of our study reveal a novel neural mechanism that allows selective attention to cause surprisingly massive rate changes in V1 and V2 neuronal populations that respond to a weak (low-contrast) target stimulus in the presence of a strong (high-contrast), nearby distracter stimulus. Target facilitation, depending on the difference between target and distracter-evoked population rates, and distracter suppression, depending on distracter-evoked rates, result in small population rate advantages in favor of the target stimulus if monkeys successfully perform a demanding shape-tracking task. It will be a challenge for future work to investigate the detailed neural mechanisms underlying these highly flexible, attention- and context-dependent firing rate modulations and their possible role in controlling selective information routing and processing in cortical networks of the visual system and beyond.

**Acknowledgments**

This work was funded by the Deutsche Forschungsgemeinschaft (DFG, German Research Foundation) - 429934733, 331514942, and 238990875. In addition, we thank Aleksandra Nadolski, Peter Bujotzek, and Katja Taylor for training and valuable technical assistance and Katrin Thoß and Ramazani Hakizimana for animal care.

Sclar, G., Maunsell, J. H. R., & Lennie, P. (1990). Coding of image contrast in central visual pathways of the macaque monkey. *Vision Research*, *30*(1), 1–10. https://doi.org/10.1016/0042-6989(90)90123-3

Sillito, A. M., & Jones, H. E. (1996). Context-dependent interactions and visual processing in V1. *Journal of Physiology Paris*, *90*(3–4), 205–209. https://doi.org/10.1016/S0928-4257(97)81424-6

Sundberg, K. A., Mitchell, J. F., & Reynolds, J. H. (2009). Spatial Attention Modulates Center-Surround Interactions in Macaque Visual Area V4. *Neuron*, *61*(6), 952–963. https://doi.org/10.1016/j.neuron.2009.02.023

Tang, R., Chen, W., & Wang, Y. (2021). Different roles of subcortical inputs in V1 responses to luminance and contrast. *European Journal of Neuroscience*, *53*(11), 3710–3726. https://doi.org/10.1111/ejn.15233

Taylor, K., Mandon, S., Freiwald, W. A., & Kreiter, A. K. (2005). Coherent oscillatory activity in monkey area v4 predicts successful allocation of attention. *Cerebral Cortex*, *15*(9), 1424–1437. https://doi.org/10.1093/cercor/bhi023

ter Wal, M., & Tiesinga, P. H. (2017). Phase difference between model cortical areas determines level of information transfer. *Frontiers in Computational Neuroscience*, *11*(February). https://doi.org/10.3389/fncom.2017.00006

Treue, S., & Martinez Trujillo, J. C. (1999). Feature-based attention influences motion processing gain in macaque visual cortex. *Nature*, *399*(6736), 575–579. https://doi.org/10.1038/21176
42

**Figures**

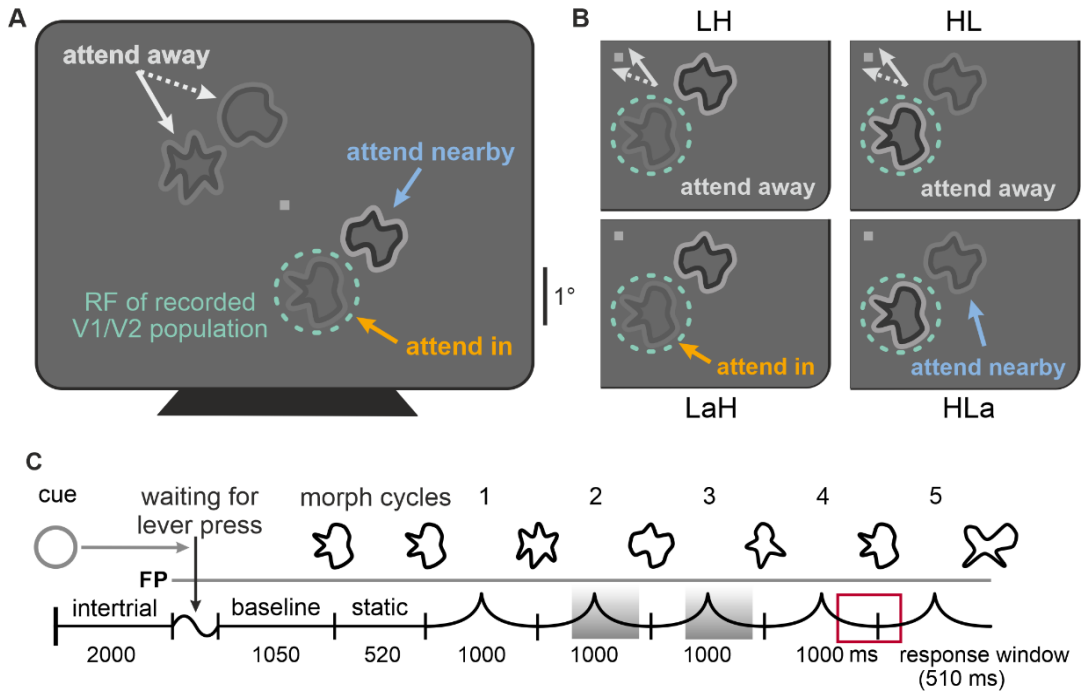

*Figure 1: Task conditions with stimulus configuration and timeline for an example trial.*
*A) Task conditions. Four complex-shaped stimuli were placed at equal eccentricity of 1.5-2º around a central fixation point. Figure displays stimulus configuration of the low-high condition where a low contrast stimulus (16%) in the lower quadrant is paired with a proximal high contrast stimulus (72%). The green dashed annulus indicates the receptive field of the recorded population of neurons. In the attend-in condition attention was directed to the receptive field stimulus. Attentional allocation to the stimulus in close vicinity to the RF is denoted as attend-nearby. Stimuli in the upper quadrant had matching luminance contrasts of 50% and could represent the target stimulus in the attend-away condition. B) Contrast settings of stimuli in lower quadrant for different task conditions. C) Timeline for target stream of an example trial with target reappearance towards the end of morph cycle 4. Position of the target stimulus was cued by faint grey annulus. The grey line represents the fixation point, which was present throughout the trial. The red box indicates the response window and the grey shaded areas the time interval over which firing rates were estimated.*

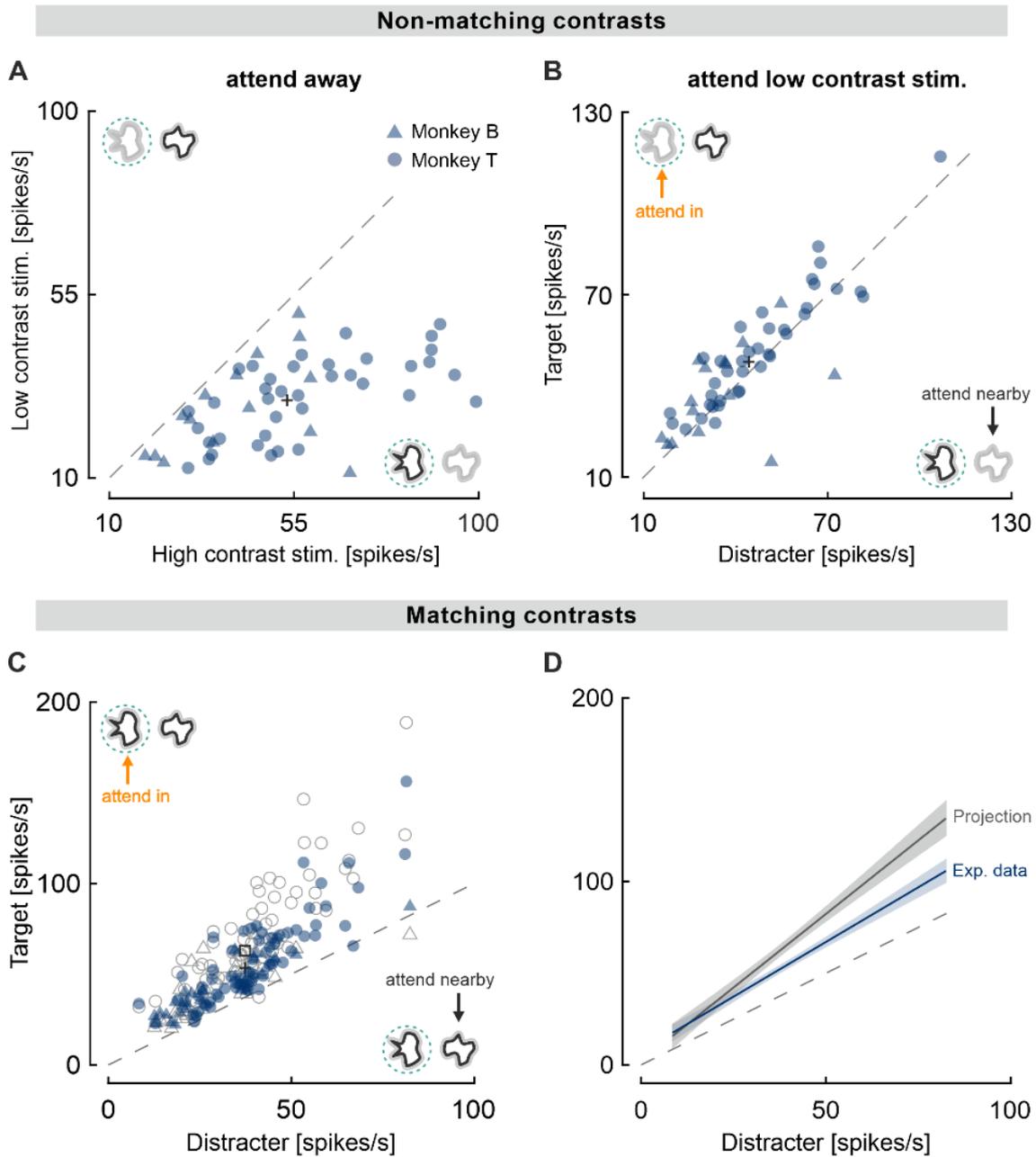

*Figure 2: Comparison of mean firing rates for competing stimuli of non-matching and matching luminance contrast.*

*Dashed lines indicate the identity line and the black cross represents the mean across all data points. **A)** Comparison of activity for low and high contrast stimuli without attention. The high contrast stimulus induces higher average MUA (mean: 53.3 spikes/s) than the adjacent low contrast stimulus (mean: 29 spikes/s) resulting in relative activity differences from 10% to 510 %. **B)** Comparison of activity for low contrast target and high contrast distracter. Attention directed to the low contrast stimulus results in higher mean firing rates (mean: 47.9 spikes/s) compared to the proximal high contrast stimulus (mean: 44.3 spikes/s). This rate advantage is robustly attained over a large activity range and for the majority of recording sites. The target stimulus has a mean rate advantage of 8.1 %. **C)** Comparison of target and distracter activity for stimulus pairings with equal contrast. Target activity exceeds the activity associated* *with the distracter stimulus in all but one cases and has a mean firing rate advantage of 43.2%. Grey symbols (projections, mean indicated by black square) mark for each session the target activity that would be reached by adding the observed rate modulations in the nonmatching condition to the distracter activity.* ***D)*** *Individual fits of the linear regression including 95% confidence bounds (shaded areas) for the actual (blue) and projected (black) data. Projection of the facilitation effects observed in the non-matching contrast configuration results in significantly higher target activity than actually observed for stimuli of matching contrast (Wilcoxon signed rank test, p = 1.87e-09), resulting in an average rate advantage of 68.6%.*

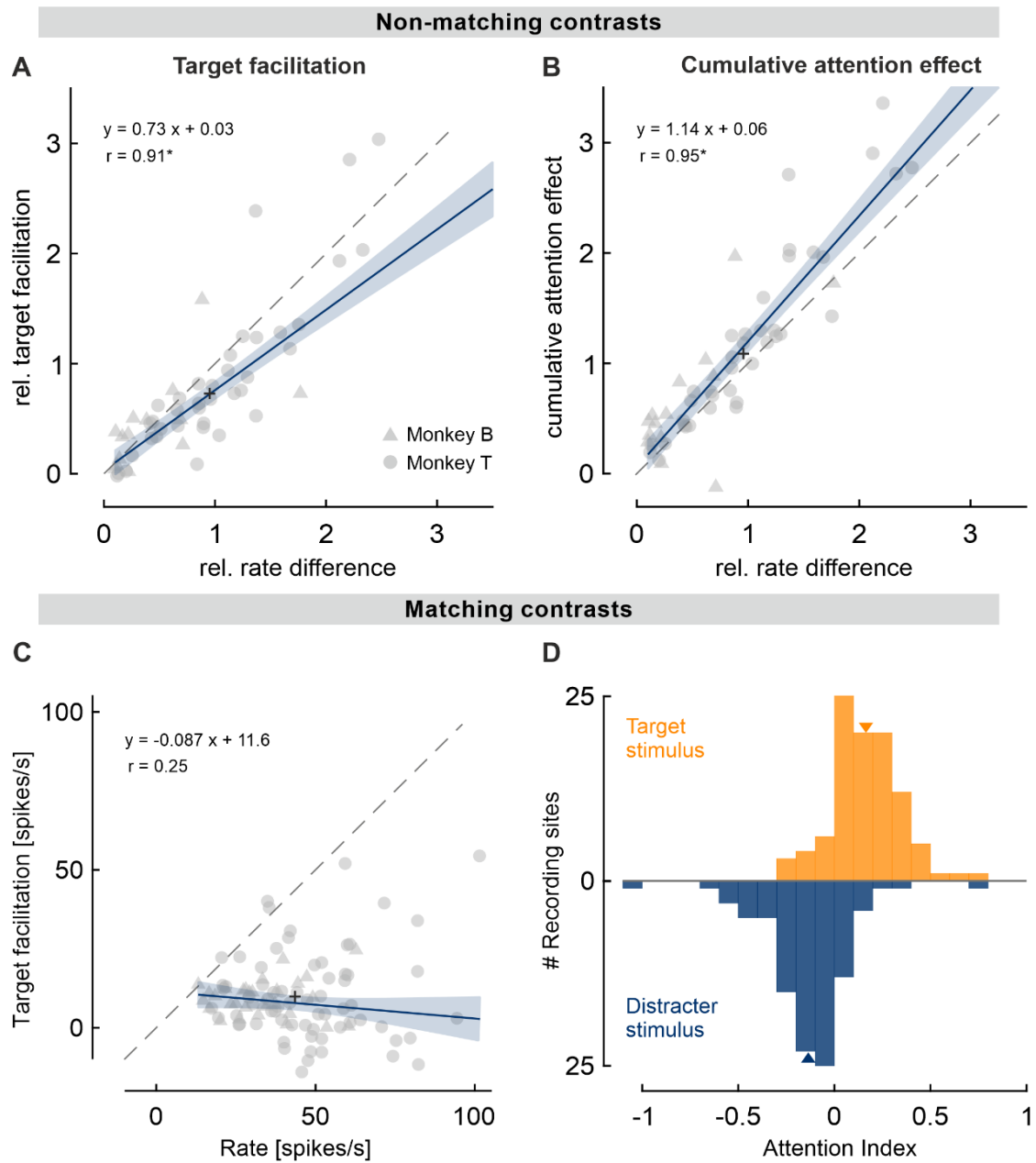

*Figure 3: Effect of attention on mean firing rates for stimulus configurations of competing stimuli with non-matching and matching contrast.*

*Dashed lines indicate the identity line, and the blue lines represent the function of the linear fit. The blue-shaded areas display 95% confidence bounds, and the black crosses indicate population means. **A-B)** results for stimulus configuration of non-matching contrasts (LH). **C-D)** results for stimulus configuration of matching contrasts (LL & HH) **A)** Relative target facilitation in dependence on the relative rate difference. Target facilitation correlates strongly with the difference in activity between the unattended low- and high-contrast stimuli (r = 0.91, p = 2.25e-21). For the mean rate difference of 83.4%, we observed an average rate gain of 65 %. The slope of the linear fit indicates that about 73% of this rate difference is compensated by target facilitation. **B)** Entire attentional modulation effect (rel. target facilitation + rel. distracter suppression) in dependence on the relative rate difference. The cumulative*

*attention effect scales by about 1.14 with the rate difference between the low and high contrast stimulus without attention and compensates for the rate difference in the majority of cases (r = 0.95, p = 3.02e-27).* ***C)*** *Target facilitation in dependence on the activity evoked by the same stimulus without attention. Changes in mean firing rates for the attended stimulus are not correlated to the purely stimulus-induced activity (r = 0.25, p = 0.13). Attention increased neuronal activity by 22.7% on average.* ***D)*** *Attentional modulation index ($R_{in} - R_{away}) \div (R_{in} + R_{away}$) for the target stimulus and ($R_{nearby} - R_{away}) \div (R_{nearby} + R_{away}$) for the distracter stimulus (n = 98). Orange and blue solid arrows represent the mean AMI for the target (0.16) and distracter (-0.14) stimulus, respectively.*

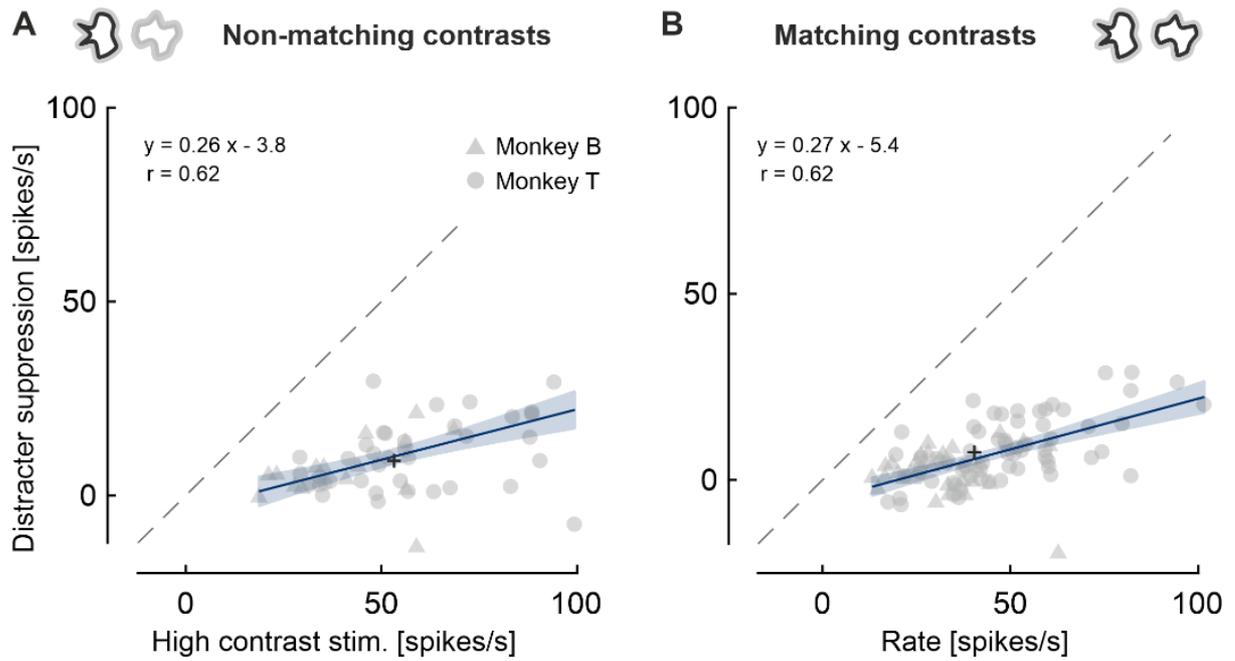

*Figure 4: Distracter suppression for competing stimuli with non-matching and matching luminance contrast.*

*Dashed lines indicate the identity line. Blue lines show the linear fit with 95% confidence bounds (shaded areas). The mean across all data points is represented by the black cross. Rate attenuation for a high contrast distracter adjacent to a low contrast target is shown in **A**), and for a high or low contrast distracter adjacent to a target of matching contrast in **B**), in dependence on the activity evoked by the same stimulus constellation without attention. Attenuation of mean firing rates for a nearby distracter correlate significantly with activity levels in the attend-away condition, in both stimulus configurations (**A**: r = 0.62, p = 1.28e-06, **B**: r = 0.62, p = 8.24e-12).*

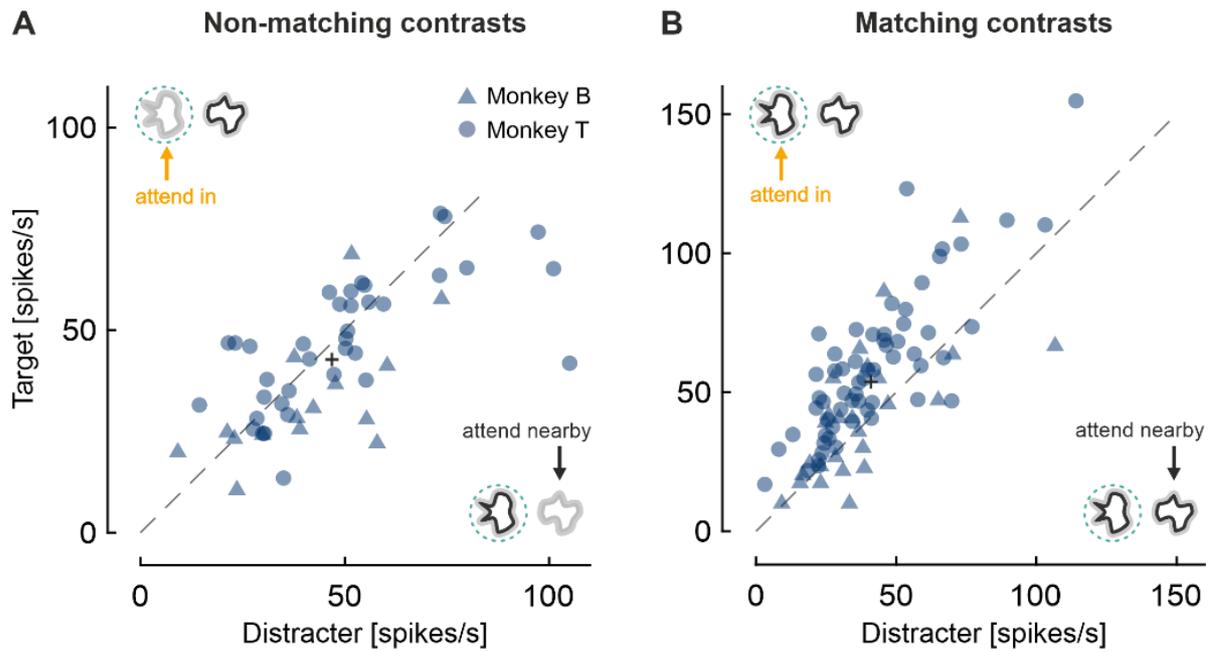

*Figure 5: Comparison of mean firing rates for target and distracter stimuli in error trials. Dashed lines indicate the identity line. **A**) Target vs. distracter activity in error trials for competing stimuli of different contrast. Activity evoked by the high contrast distracter has a mean population activity of 46.8 spikes/s, whereas the mean activity associated with the low contrast target is 42.7 spikes/s (n = 52). **B**) Target vs. distracter activity in error trials for competing stimuli of equal contrast. A mean activity for the distracter stimulus of 42.5 spikes/s is opposed by a mean activity for the target stimulus of 55.2 spikes/s (n = 80)*

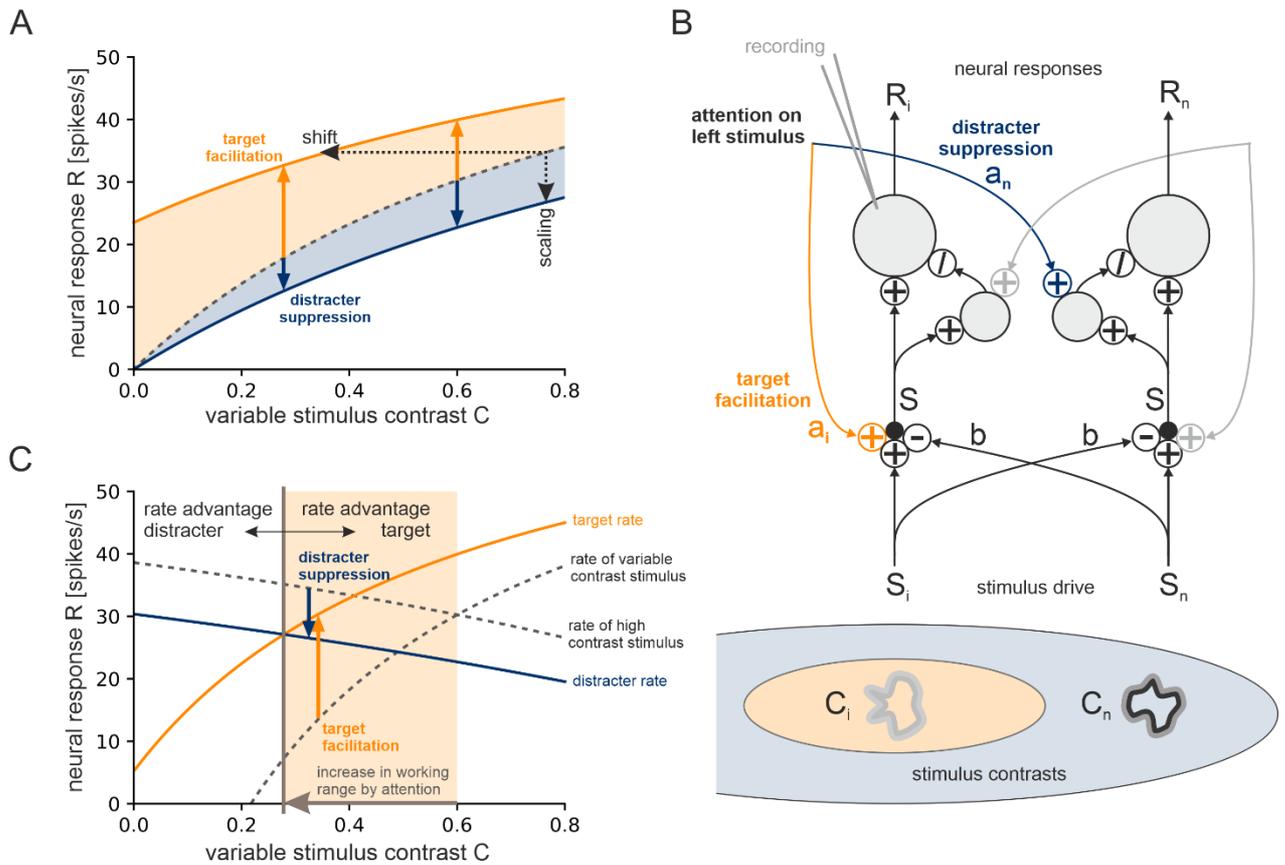

*Figure 7: Modeling response modulation by attention and stimulus configuration. A) Illustration of the neural response changes (orange and blue arrows) induced by shifting a nonlinear contrast-response function $g(S)$ (dashed line) to the left, e.g. by increasing $a_i$ in Eq.(1) (contrast gain enhancement, orange line and shaded area), or by downscaling the output via increasing $a_n$ in Eq.(1) (blue line and shaded area). While a shift induces a response increase which becomes smaller for high stimulus contrasts (as observed for target facilitation), scaling the output induces a response decrease which becomes larger for high stimulus contrasts (as observed for distracter suppression). B) Model circuit for a stimulus configuration with two nearby stimuli. Here we show a situation (LaH, cf. Fig. 1B) where a low-contrast target stimulus (bottom left) which is inside the RF of the recorded neuron (gray electrode, top) was paired with a high-contrast distracter (bottom right). Variables are as defined in the text and in the equations for $S_I, S_N, R_i$ and $R_n$. Target facilitation is indicated by the orange arrows and region in the visual field, distracter suppression by the blue arrows and region. C) Increase in working range (grey arrow and yellow region) induced by the combined effects of target facilitation and distracter suppression. The two dashed lines show the neural responses to a stimulus configuration with a high-contrast stimulus ($C_n = 60\%$) paired with a nearby stimulus of variable contrast (indicated on the horizontal axis) -- a rate advantage for the variable contrast stimulus is only possible if its contrast is higher than that of the variable contrast stimulus. The orange and blue lines show the neural responses emerging when the variable contrast stimulus becomes the target -- a rate advantage is now established for much lower target contrasts and results from a combination of target facilitation (orange arrow) and*

*distracter suppression (blue arrow). Parameters for the graphs in (**A**) and (**B**) were taken from the fits to population averages for monkey T.*

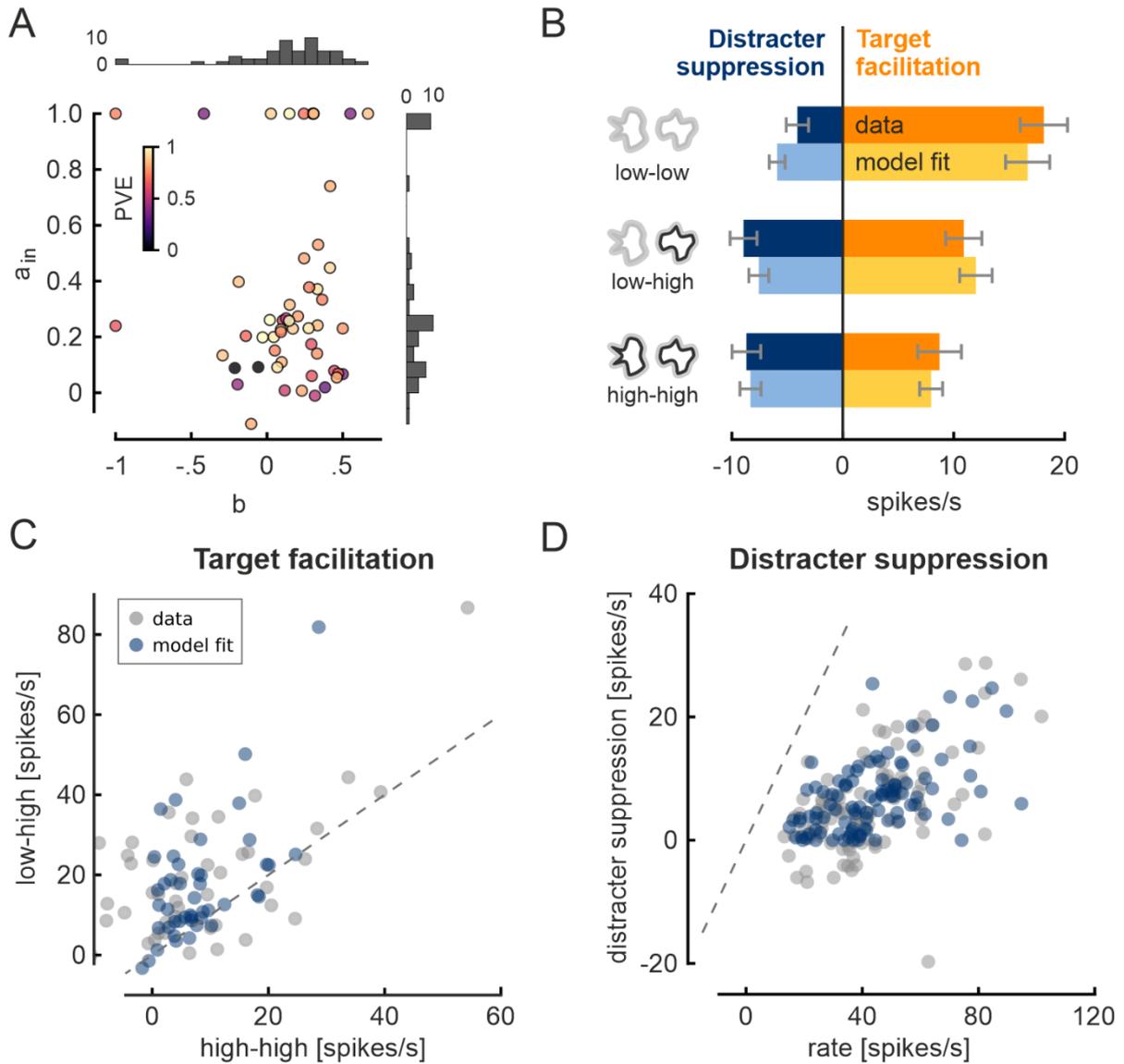

*Figure 8: Model fits for individual sites and comparison with experimental data. A) Joint distribution of fit parameters for attention parameter $a_i$ and surround suppression parameter $b$. The bar graphs on the top and on the right show the corresponding marginal distributions over $a_i$ and $b$, respectively. Quality of the individual fits is indicated by the colour scale inset. B) Effect sizes of target facilitation and distracter suppression for the experimental data (blue) and the model fits (orange) in the low-high, low-low and high-high configurations. C) Target facilitation in presence of a nearby high-contrast distracter, plotted for low target contrasts (vertical axis) versus high target contrasts (horizontal axis). D) Distracter suppression of a stimulus induced by attention placed on a nearby stimulus with matching contrast.*